\documentclass[letterpaper,prd,twocolumn,unsortedaddress]{revtex4}[10pt]
\usepackage{graphicx}
\usepackage{epsfig}
\usepackage{hyperref}

\begin{document}
\newcommand{\beq}{\begin{eqnarray}}
\newcommand{\eeq}{\end{eqnarray}}
\newcommand{\nn}{\nonumber}
\def\ltap{\ \raise.3ex\hbox{$<$\kern-.75em\lower1ex\hbox{$\sim$}}\ }
\def\gtap{\ \raise.3ex\hbox{$>$\kern-.75em\lower1ex\hbox{$\sim$}}\ }
\def\CO{{\cal O}}
\def\CL{{\cal L}}
\def\CM{{\cal M}}
\def\tr{{\rm\ Tr}}
\def\CO{{\cal O}}
\def\CL{{\cal L}}
\def\CM{{\cal M}}
\def\mpl{M_{\rm Pl}}
\newcommand{\bel}[1]{\be\label{#1}}
\def\al{\alpha}
\def\bt{\beta}
\def\eps{\epsilon}
\def\eg{{\it e.g.}}
\def\ie{{\it i.e.}}
\def\mn{{\mu\nu}}
\newcommand{\rep}[1]{{\bf #1}}
\def\be{\begin{equation}}
\def\ee{\end{equation}}
\def\bea{\begin{eqnarray}}
\def\eea{\end{eqnarray}}
\newcommand{\eref}[1]{(\ref{#1})}
\newcommand{\Eref}[1]{Eq.~(\ref{#1})}
\newcommand{\gsim}{ \mathop{}_{\textstyle \sim}^{\textstyle >} }
\newcommand{\lsim}{ \mathop{}_{\textstyle \sim}^{\textstyle <} }
\newcommand{\vev}[1]{ \left\langle {#1} \right\rangle }
\newcommand{\bra}[1]{ \langle {#1} | }
\newcommand{\ket}[1]{ | {#1} \rangle }
\newcommand{\ev}{{\rm eV}}
\newcommand{\kev}{{\rm keV}}
\newcommand{\Mev}{{\rm MeV}}
\newcommand{\gev}{{\rm GeV}}
\newcommand{\tev}{{\rm TeV}}
\newcommand{\mev}{{\rm MeV}}
\newcommand{\meV}{{\rm meV}}
\newcommand{\mnu}{\ensuremath{m_\nu}}
\newcommand{\nnu}{\ensuremath{n_\nu}}
\newcommand{\mlr}{\ensuremath{m_{lr}}}
\newcommand{\acc}{\ensuremath{{\cal A}}}
\newcommand{\mav}{MaVaNs}
\newcommand{\disc}[1]{{\bf #1}} 
\newcommand{\mh}{{m_h}}
\newcommand{\hb}{{\cal \bar H}}
\newcommand{\me}{\mbox{${\rm \not\! E}$}}
\newcommand{\met}{\mbox{${\rm \not\! E}_{\rm T}$}}
\newcommand{\MPl}{M_{\rm Pl}}
\newcommand{\ra}{\rightarrow}
\pagestyle{plain} 
\title{``Light from chaos" in two dimensions}
\author{Erich Poppitz	}
\email{poppitz@physics.utoronto.ca}
\author{Yanwen Shang}
\email{ywshang@physics.utoronto.ca}
\affiliation{Department of Physics, University of Toronto, 60 St George St.,
Toronto, Ontario M5S~1A7,  Canada}

\begin{abstract}
We perform a Monte-Carlo study of the lattice   two-dimensional gauged XY-model. Our results confirm 
the strong-coupling expansion arguments that for sufficiently small values of the spin-spin coupling the 
``gauge symmetry breaking" terms decouple and the long-distance physics is that of the unbroken  pure 
gauge theory. We  find no evidence for the existence, conjectured earlier, of massless states near a 
critical value of the spin-spin coupling.
We  comment on recent remarks in the literature on the use of  gauged XY-models in 
proposed  constructions of chiral lattice gauge theories. 
\end{abstract}

\maketitle
\twocolumngrid


\section{Motivation and summary} 

The gauged two-dimensional lattice XY-model has been studied for quite some time. Its simplicity allows 
for analytic studies via lattice dualities and strong-coupling expansions \cite{Peskin:1977kp}  as well as 
for numerical analysis, in either a Hamiltonian \cite{Jones:1978yr} or Euclidean (see, e.g., 
\cite{Grunewald:1986ib}) formulation.  The model exhibits many phenomena found in more complicated 
higher-dimensional theories, notably the presence of both confining and Higgs phases 
\cite{Peskin:1977kp, Jones:1978yr, Fradkin:1978dv, Banks:1979fi}.

The motivation for this short study stems from our interest in using the gauged XY-model, as well as  its 
nonabelian higher-dimensional analogues \cite{Fradkin:1978dv,Forster:1980dg, Lang:1981qg},  in  lattice 
constructions of chiral gauge theories. The lattice formulation of chiral gauge theories is an outstanding 
problem with no practical solution yet, despite much recent progress; for reviews, see 
\cite{Golterman:2000hr}. While not our  topic here, we note that the recent constructions of 
\cite{Bhattacharya:2006dc, Giedt:2007qg}  have some attractive features (for an earlier  
proposal of similar flavor, see \cite{Creutz:1996xc}).
Most importantly,  they may offer a  way around the difficult and unsolved problem of the explicit 
nonperturbative construction of the fermion measure for general chiral lattice  gauge theories. 
Ref.~\cite{Bhattacharya:2006dc} aims to achieve this by combining older ideas \cite{Eichten:1985ft} to 
use non-gauge strong dynamics  to decouple the mirror fermions in a vectorlike theory  with the recently 
discovered  exact lattice chirality of the Neuberger-Dirac operator.  

A   concrete two-dimensional  realization   of the proposal \cite{Bhattacharya:2006dc} has  the gauged 
XY-model as an essential part and was studied in various limits in  \cite{Giedt:2007qg}. Recently, 
\cite{Suzuki:2007zf}
 considered the perturbative spectrum of the models of  \cite{Bhattacharya:2006dc} at nonzero gauge 
coupling. It was claimed there that the gauge boson is always massive and, therefore, the construction of  
\cite{Bhattacharya:2006dc,Giedt:2007qg}  was argued to be irrelevant to the study of unbroken chiral 
gauge theories, on account of this fact alone. 

This brings us to the main topic of this paper. The arguments of  \cite{Suzuki:2007zf} do not invoke either 
the fermions or the   strong mirror dynamics (admittedly, complicated and not yet completely understood), 
but concern only   the spectrum of the gauged XY-model,  arguing that it only has a  phase with a 
massive gauge boson.
 This conclusion contradicts strong-coupling expansion arguments   for the decoupling of the 
 ``gauge-breaking" terms for small values of the spin-spin coupling. As these arguments have been  made many 
times in the past \cite{Jones:1978yr, Fradkin:1978dv, Forster:1980dg},  we will not repeat them at length here. 
 
 Instead, given that confusion around the issue appears to still persist,  we use Monte-Carlo 
simulations to argue the same point.  These methods allow us to also  study  regions beyond the reach of   
strong-coupling expansions or  perturbation theory and to  observe in  detail how the infrared physics 
changes smoothly as  the   XY-model spin-spin coupling $\kappa$ is varied.
We  show that for $\kappa < \kappa_c$, with $\kappa_c = {\cal{O}}(1)$, the long-distance physics of the 
gauged XY-model is that of the unbroken pure gauge theory. We numerically demonstrate that, in accord 
with  the strong-coupling expansion, for subcritical  $\kappa$ the leading effect of the spin-spin coupling 
can be incorporated into a  shift of the gauge coupling. 

We also study the spectrum in coupling regimes inaccessible by controlled analytic methods. The lattice 
Hamiltonian of the gauged XY-model was studied long ago  via  a strong-coupling expansion in the gauge 
coupling, valid for arbitrary values of $\kappa$, with a subsequent Pad\' e resummation to small gauge 
coupling \cite{Jones:1978yr}.  A dip in the spectrum of charged particle pairs near a critical value of $
\kappa$  of order unity  was found, and it was conjectured  that it may indicate the presence of massless 
states.  We look for such states by measuring the susceptibilities of the relevant operators. We  find a 
slight increase of the susceptibility  around similar values of $\kappa$, consistent with the results of the 
Pad\' e resummation of the strong-coupling series. However,  we find no finite-size scaling of the 
susceptibility and hence no evidence for  massless states in the intermediate ``critical" regime.  
 
Finally, as far as  lattice formulations of chiral gauge theories are concerned, 
our results address the objection of \cite{Suzuki:2007zf}  to the proposal \cite{Bhattacharya:2006dc}.  
Much remains to be done to see whether   constructions along the lines of  \cite{Creutz:1996xc},
\cite{Bhattacharya:2006dc} fulfill their designed goal---to give rise to long-distance gauge theories with  
chiral fermions in complex representations. While we hope to report more on at least some of the  outstanding issues  in 
the future, our main point  here is that the reason for their failure would have to be more subtle.

\section{The model}
\label{themodel}

The Euclidean action of the gauged XY-model is:
\beq
\label{XYaction}
- S = \sum_{x} \left({\beta \over 2}  \prod\limits_{plaq} U   + {\kappa \over 2} \sum\limits_{\hat{\mu}}  
\phi^*_x U_{x, \hat{\mu}} \phi_{x+\hat{\mu}} \right)+ {\rm h.c.} ,
\eeq
where $U_{x, \hat{\mu}} = e^{i A_{x, \mu}}$ is the  link group element,$ \prod\limits_{plaq} U$ is the usual 
plaquette Wilson action,  and $\phi_x = e^{i \eta_x}$ is a unitary Higgs field; $x$ denotes lattice sites on a 
two-dimensional square lattice and $\hat{\mu}$ is a unit lattice vector in the $\mu = 1,2$ direction.
Here $A_{x,\mu}$ and $\eta_x$ are angular variables defined on links and sites, respectively; throughout the paper the lattice spacing is set to unity.

The partition function of the model is defined via a lattice path integral over all angular variables.  The 
action can be more succinctly expressed (and the simulation simplified) if one notes that upon a change 
of the link variables the dependence of the action on $\eta_x$ can be completely eliminated, leaving us 
with a simpler theory to study:
\beq
\label{XYaction2}
- S = \sum_{x} \left( \beta \cos F_{x} + \kappa \sum_{\hat{\mu}} \cos A_{x, \mu} \right)~,
\eeq
where $F_{x} = A_{x + 1, 2} -  A_{x, 2} - A_{x+2,1} + A_{x, 1}$ is the lattice gauge field strength. In the 
naive continuum limit, taking $\beta ={1\over  g^2}$, the action (\ref{XYaction2}) becomes: 
\beq
\label{XYaction3}
&&-S_{naive}  = \\
&&  \sum_x   -{1 \over 2 g^2}\; F_{x}^2\; - {\kappa \over 2} (A_{x,1}^2 + A_{x,2}^2)   + \kappa\;  {\cal{O}}
(A^4)  ,\nonumber
\eeq
and describes a massive gauge boson of mass $m_W^2 = g^2  \kappa$---a conclusion which  holds for sufficiently large $\kappa$. For small $\kappa$, however, the perturbative expansion around the  
naive continuum limit is not a good guide to the  spectrum of the theory, because the naive 
continuum limit misses two important points: the angular nature of the variables and  the fact that at small 
$\kappa$ the quantum fluctuations of the longitudinal component of the gauge field $A_{x, \mu}$ are not 
suppressed. For $\kappa \ll 1$, the ${\cal{O}}(A^4)$ terms lead to strong interactions 
between the longitudinal component of $A_\mu$---the fluctuations of these modes are not suppressed 
by  the gauge invariant kinetic term and  for small $\kappa$ the spin-spin (or ``mass") term does not suppress them either. In contrast, for 
$\kappa \gg 1$, these modes' fluctuations are   suppressed by the ``mass" term in 
(\ref{XYaction3}) and the $(A_\mu)^4$ interactions are irrelevant. 

For small   $\kappa$, therefore, a strong coupling expansion in $\kappa$ is  more  appropriate than the 
perturbative expansion---the longitudinal components of the gauge field   strongly fluctuate on the scale 
of the lattice spacing,
are thus heavy, and can be integrated out. The small-$\kappa$ expansion is
 most conveniently performed in the original gauge invariant formulation in terms of the  variables $A_{x,
\mu}$ and $\eta_x$ of (\ref{XYaction}). To leading order in $\kappa$, the effect on the infrared physics of 
the gauge field  of   the rapid fluctuations of the Higgs field $\phi_x$ is to renormalize the gauge coupling 
\cite{Fradkin:1978dv,Forster:1980dg}. 
 The long-distance---compared to the correlation length of the Higgs field excitations,  which is smaller 
than the lattice spacing for $\kappa < \kappa_c$---effective action is of the form:
 \beq
 \label{XYaction4}
 -S_{\kappa < \kappa_c} \simeq  \sum_{x}  \left(\beta + {\kappa^4\over 8}\right) \cos F_{x} + \ldots
 \eeq
 where dots denote higher-dimensional gauge invariant terms.  Computing the coefficient of the 
 leading-order shift of the bare gauge coupling in (\ref{XYaction4}), as well as those of the higher-order  terms 
(leading to larger plaquettes and  omitted in (\ref{XYaction4})), is trivially done by using the small-$\kappa
$ expansion of the partition function defined by    (\ref{XYaction}), and by integrating over the fluctuations 
of the compact scalar field $\phi_x$. 
 
 Equivalently, we show in the next Section that the coefficient of $\kappa^4$ in (\ref{XYaction4}) can be 
read off the results of our simulation (clearly, this is numerically possible only if  $\beta$ is not too large). 
The Monte-Carlo methods  also allow us to see how the infrared physics changes as a function of $
\kappa$ and to study the spectrum of various operators in regimes where neither the strong-coupling 
expansion nor perturbation theory apply.

\section{The Monte-Carlo study}

Our Monte-Carlo study focuses on the evaluation of several quantities: the Wilson loop, the Polyakov 
loop, and several zero-momentum correlators, or susceptibilities. We study their variation with $\kappa$, 
$\beta$, and the lattice size $N$, our primary interest being the $\kappa$ dependence for small  
gauge coupling. 

This choice of observables is motivated by  the nature of two-dimensional physics: 
we remind the reader that the two-dimensional pure $U(1)$  gauge theory (which we argue describes the 
infrared of the gauged XY-model for subcritical $\kappa$) has no local propagating degrees of freedom. 
Nonetheless, it  exhibits nontrivial  dynamics manifesting itself in a nonzero expectation value of the 
Wilson loop, with an area law behavior for all values of the gauge coupling. Similarly, the Polyakov loop 
expectation value vanishes in the pure gauge theory for any coupling and temperature. On the other 
hand, the gauged XY-model  with nonzero sufficiently large $\kappa$  has propagating local degrees of 
freedom, and information on their spectrum can be inferred from correlators of local operators.  

All our simulations are performed via the Metropolis algorithm, using the partition function in ``unitary 
gauge"  defined by the action (\ref{XYaction2}). The autocorrelation time for the various observables is 
monitored. The  largest autocorrelation time---that of the Polyakov loop, varying between about ten and a 
few hundred of lattice updates, depending on $\beta$ and $\kappa$---is used to estimate the errors. 

\subsection{The Wilson and Polyakov loops }

We denote by $W_{ij}$ the expectation value of the   Wilson loop operator for a rectangular  loop of size   
$i \times j$. 
 In the pure gauge theory, $\kappa = 0$, the Wilson loop expectation value can be exactly calculated. The 
string tension $\sigma$  of the pure $U(1)$ gauge theory, which confines for all values of the gauge 
coupling,  see e.g. \cite{Gross:1980he}, is easily shown to be, in the infinite volume limit:
\beq
\label{wilsongauge}
\sigma(\beta) = - {\ln W_{ij} \over i \times j} = - \ln { I_1(\beta) \over I_0(\beta)}~ , 
\eeq
where $I_{1}, I_0$ are Bessel functions. At strong coupling, the string tension is $\sigma =  \ln 4 g^2 +
{\cal{O}}(1/g^2)$, while at  weak coupling one instead obtains  $\sigma \simeq   g^2$. 

Finite-volume corrections to the string tension will sometimes be significant in our  simulations. In order to 
disentangle these effects, we have computed the Wilson loop expectation value in the pure gauge theory  
in finite volume, using the dual representation (see, e.g., the second reference in \cite{Peskin:1977kp}) of the partition function. We thus obtain   for $W_{ij}$ in finite volume:
\beq
\label{wilsongaugevolume}
W_{ij} =  { \sum\limits_{n} I_n (\beta)^{N^2 - A} I_{n-1}(\beta)^{A}  \over \sum\limits_{m} I_m 
(\beta)^{N^2} }~,
\eeq
where $A = i j$ is the area of the loop, $N^2$ is the lattice volume, and the sums are over all integers. It 
is easy to check a few simple limits: as $N\rightarrow \infty$     (\ref{wilsongaugevolume}) reproduces the 
result  (\ref{wilsongauge}) for the string tension; also, it  correctly yields $W_{NN} = 1$. We  use 
(\ref{wilsongaugevolume}) to show that for subcritical $\kappa$  our measured Wilson loop expectation 
values  reduce to the ones  of the pure gauge theory  also in cases where finite-size effects are 
numerically significant.

Turning on $\kappa \ne 0$,  for large $\kappa$ we expect from  Eqn.~(\ref{XYaction3})   to obtain a 
perimeter law for the Wilson loop. For small $\kappa$, from Eqn.~(\ref{XYaction4}) we should expect an 
area law instead. To avoid confusion, we stress that  the theory does have charged matter excitations---
the quanta of the Higgs field $\phi_x$---which can screen external charges and lead to perimeter law for 
the Wilson loop. In the small-$\kappa$ limit, however,  the charged excitations are heavier than the UV 
cutoff. While for any charged fields' mass a sufficiently long ``string" will eventually find it energetically  
favourable to break, we can imagine taking the continuum and large-volume limits so that the breaking 
does not occur even for the longest possible string, i.e., $\sigma  N/2  < m_S$, where $N$ is the lattice 
size and $m_S$ is the lowest energy  of a pair of charged particles, much higher than the UV cutoff 
($m_S \sim \kappa^{-1}$ in the strong-coupling expansion \cite{Jones:1978yr}).

\begin{figure}
\includegraphics[width=3.4in]{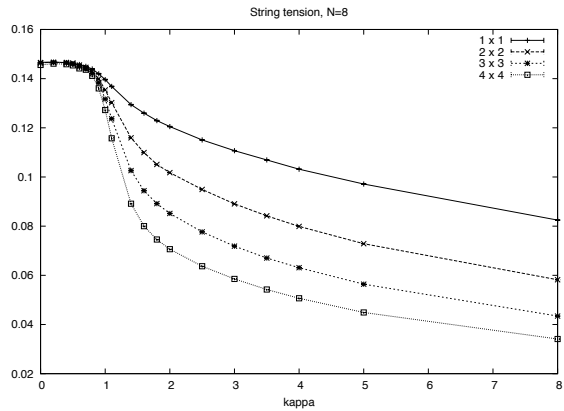}
\caption{The dependence of $|\ln W_{ij}|/(ij)$ on $\kappa$ for $i=j=1,2,3,4$, for an $8^2$ lattice with $
\beta=4$. The figure clearly shows a transition to an area law at   $\kappa < 0.5$, with string tension 
given 
by the expression (\ref{wilsongauge}) for the pure gauge theory,  $\sigma(4) \approx .147$. A perimeter 
law is approached at large $\kappa$.}
\label{fig:wilsonareaperimeter8} 
\end{figure}

\begin{figure}
\includegraphics[width=3.4in]{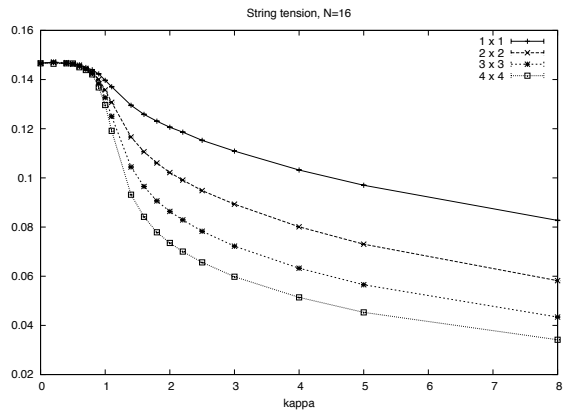}
\caption{Same as Fig.~\ref{fig:wilsonareaperimeter8}, but  for a $16^2$ lattice with $\beta=4$. }
\label{fig:wilsonareaperimeter16} 
\end{figure}

\begin{figure}
\includegraphics[width=3.4in]{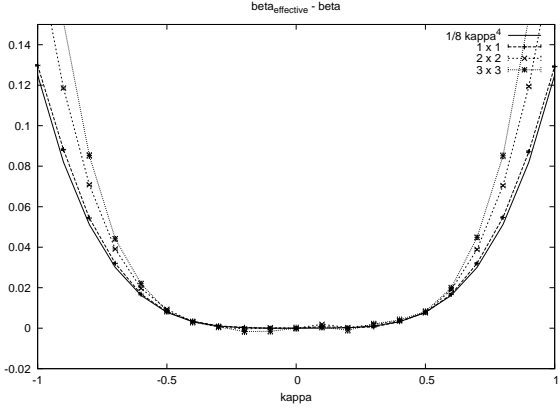}
\caption{The fit of $\Delta\beta(\kappa)$ vs. $\kappa^4/8$, see Eqn.~(\ref{XYaction4}) for the Wilson loop expectation value. Fit uses 
an $8^2$ lattice, $1\times 1$, $2 \times2$ and $3\times 3$ loops.}
\label{fig:wilsonkappa} 
\end{figure}

We begin the discussion of our Monte-Carlo results by showing the dependence of the ``string tension" $
\sigma_{ij}$=$-\ln W_{ij}/(ij)$ on $\kappa$ for Wilson loops with $i$=$j$=$1, 2, 3, 4$. 
On Fig.~\ref{fig:wilsonareaperimeter8}, we show $\sigma$ for an $8^2$ lattice and $\beta=4$, while on 
Fig.~\ref{fig:wilsonareaperimeter16}, we show $\sigma$ on a $16^2$ lattice for the same $
\beta=4$.  For these loop sizes and values of $\beta$ the finite volume effects are small.  For small $
\kappa < {\cal{O}}(1)$, we see that in both cases, the pure gauge theory value of the string tension  $
\sigma(4)\simeq .147$, as given by Eqn.~(\ref{wilsongauge}), is reproduced. 
 For large values of $\kappa \gg 1$, on the other hand, we see a gradual transition to a perimeter law for 
the Wilson loop, with the ``string tensions" for $\kappa = 4, 5, 8$ scaling as inverse powers of $\sqrt{A}$
($\kappa = 8$ comes closest to pure perimeter law).

 The strong-coupling expansion Eqn.~(\ref{XYaction4}), predicts that  for sufficiently small $\kappa$ the  
Wilson loop expectation value should be given by the same equation as in the pure 
gauge theory, Eqn.~(\ref{wilsongauge}), but with $\beta \rightarrow \beta + \kappa^4/8$, so long as the loop is 
larger than the correlation length of the massive charged excitations.  The small effect due to the $
{\cal{O}}(\kappa^4)$ shift of the bare gauge coupling   can not be seen for the  relatively large values of $
\beta$ on Figs.~\ref{fig:wilsonareaperimeter8}, \ref{fig:wilsonareaperimeter16}. 
To test this prediction and see for what values of $\kappa$ it breaks down,  we have measured $W_{11}$, 
$W_{22}$, and $W_{33}$ as a function of $\beta$, near $\beta \simeq 1.6$,  for a number of values of $
\kappa$, ranging from $-1$ to $1$ with spacing $\delta\kappa  = 0.1$. Then, we determine the function 
$\beta(\kappa)$ such that $W_{ii}$ is constant. On Fig.~\ref{fig:wilsonkappa}, we plot $\Delta\beta =  \beta(\kappa) - 
\beta$ vs. the small-$\kappa$ prediction of the strong-coupling expansion, $\Delta\beta_{str.coupl.} 
\simeq \kappa^4/8$. The agreement for $\kappa \le 0.5$ is quite persuasive.

\begin{figure}
\includegraphics[width=3.4in]{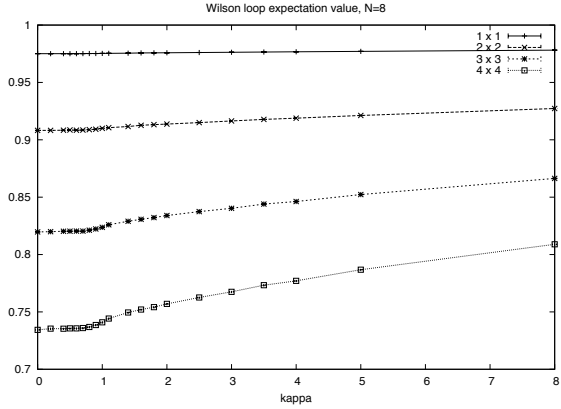}
\caption{Expectation values of Wilson loops of sizes $1^2, 2^2, 3^2, 4^2$, for $\beta = 20$, for an $8^2$ 
lattice as a function of $\kappa$. For $\kappa \le 0.5$ the results are  in agreement with the finite-size 
formula (\ref{wilsongaugevolume}) for the pure gauge theory, which  gives: $W_{11} \approx .975$, 
$W_{22}\approx .908$, $W_{33}\approx.819$, and $W_{44}\approx.732$. }
\label{fig:wilson8} 
\end{figure}

We present one more plot  of the Wilson loop expectation value, this time for the larger value of $\beta = 
20$ and an $8^2$ lattice,
on Fig.~\ref{fig:wilson8}, where we show $W_{ii}$ of sizes $1^2$, $2^2$, $3^2$, and $4^2$,  as function of $\kappa$.   
The prediction for $W_{ii}$ of the pure gauge theory,   including  finite-size effects, 
Eqn.~(\ref{wilsongaugevolume}), is given in the caption of Fig.~\ref{fig:wilson8}; for  $\kappa<0.7$, the agreement with the numerical results is evident.

\begin{figure}
\includegraphics[width=3.4in]{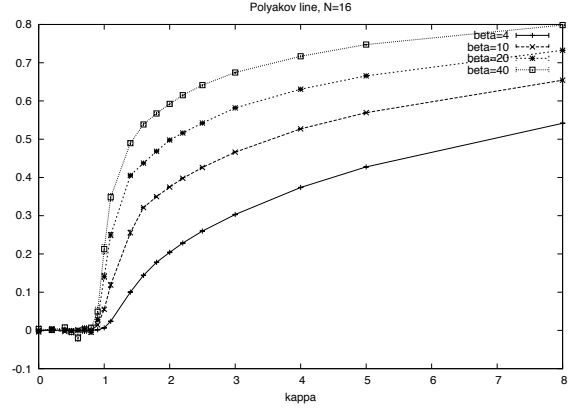}
\caption{Expectation value of the Polyakov loop on a  $16^2$ lattice as a function of $\kappa$. A 
transition to a confining phase for $\kappa \le 1$ is clearly visible.  }
\label{fig:polyakovloop} 
\end{figure}

 Finally, we study the expectation value of the  Polyakov loop operator around  the $\mu$-th compact 
direction, $P_\mu$; we recall that its vanishing signals the onset of a confining phase where the free 
energy of a isolated quark is infinite. On Fig.~\ref{fig:polyakovloop}, we show $P_\mu$ (the loops in the 
two directions are identical, up to numerical errors) measured on a $16^2$ lattice, for a range of values of 
$\beta$, as a function of $\kappa$. As with the Wilson loop, the fact that   the Polyakov loop achieves the 
value  appropriate to the pure gauge theory 
 (zero)    at $\kappa < 1$ is clearly visible on the figure.

\subsection{Zero-momentum Green's functions}

In this Section, 
 we study the following zero-momentum correlators (susceptibilities):
\beq
\label{susc}
\chi_{\mu \nu}^S = \sum\limits_x \langle \cos A_{x, \mu} \cos A_{0, \nu} \rangle^C~, \nonumber \\
\chi_{\mu \nu}^V = \sum\limits_x \langle \sin A_{x, \mu} \sin A_{0, \nu} \rangle^C~,
\eeq
where   $\langle ...\rangle^C$ denotes a connected correlator; in this Section, we take the indices to be $\mu,\nu  = 0,1$. 

We use the  susceptibilities to look for massless states in the relevant scalar ($S$) and vector ($V$) 
channel via a large-volume scaling. 
In a Hamiltonian formulation, the  operator Re$(\phi_x^* U_{x, \hat{1}} \phi_{x + \hat{1}})$, equal to $\cos 
A_{x,1}$ in unitary gauge, creates pairs of oppositely charged particles, while ${\rm Im}( \phi_x^* U_{x, 
\hat{1}} \phi_{x + \hat{1}}) = \sin A_{x,1} $ creates (massive) vector states. The spectrum of the lattice 
gauged XY-model Hamiltonian was studied in ref.~\cite{Jones:1978yr} via an expansion in $1/g^2$, valid 
 for all values of $\kappa$. Subsequently, a Pad\' e resummation was 
 used to  continue the series to $g^2 \rightarrow 0$.  The $[4,4]$ Pad\' e approximant of the scalar 
correlator 
 computed there showed  a tendency for  scalar (S) states to become light for  $\kappa$ near $
\kappa_c$ (see Figs.~8, 9 of ref.~\cite{Jones:1978yr}).  
 While ref.~\cite{Jones:1978yr} admitted that   it is not clear that the procedure used to arrive at this result 
is to be trusted,  it conjectured that  there  may be massless states  near $\kappa_c$.

 \begin{figure}
\includegraphics[width=3.4in]{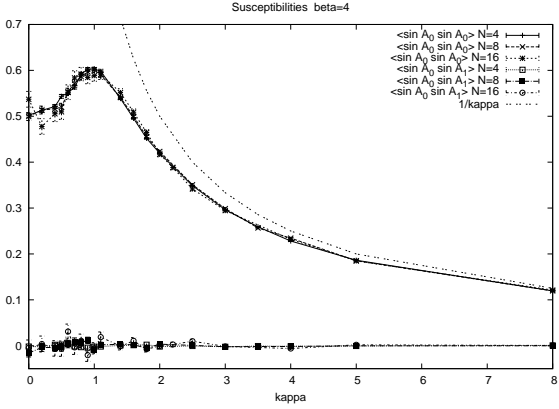}
\caption{The  values of the vector susceptibilities $\chi_{00}^V$ (upper set of curves) and $\chi_{01}^V$  
(vanishing lower curves) as functions of $\kappa$ for lattices of size  $4^2, 8^2, 16^2$. The $1/\kappa$ 
behavior expected in perturbation theory, see Eqn.~(\ref{mW}), valid for large $\kappa$, is also 
indicated.}
\label{fig:susc_sin} 
\end{figure}

 \begin{figure}
\includegraphics[width=3.4in]{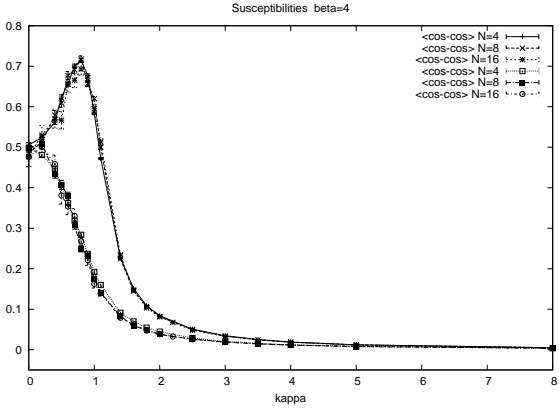} 
\caption{The two eigenvalues of the scalar susceptibility matrix $\chi_{\mu\nu}^S$ as functions of $\kappa
$ for lattices of size $4^2, 8^2, 16^2$. Clearly, no large-$N$ scaling is observed near the transition.  }
\label{fig:susc_cos}
\end{figure}

We first show our results for the  vector, $\langle \sin A_\mu \sin A_\nu \rangle$, susceptibilities 
(\ref{susc})  on Fig.~\ref{fig:susc_sin}, for lattices of various sizes and for $\beta =4$.
A few remarks are in order. First,  the off-diagonal ($\mu\nu = 01$)  components of the vector correlators 
vanish due to the $A_0 \rightarrow   A_1$, $A_1 \rightarrow - A_0$ symmetry. Second, the $00$ and 
$11$ components are identical due to the same $Z_4$ rotational symmetry of the Euclidean action; 
hence, we only show the $00$ component. Third, it is clear from the figure that there is no large-volume 
scaling of the vector susceptibility for any value of $\kappa$, and hence no indication that massless 
states are present; we have also checked that this lack of scaling persists for larger values of $\beta$. 
Finally, as explained in Section \ref{themodel}, for  $\kappa \rightarrow \infty$  perturbation theory is valid 
and the susceptibility (\ref{susc}) can be approximated by: 
\beq
\label{mW}
\chi_{00}^V \simeq \sum\limits_x \langle A_{x, 0} A_{0,0} \rangle^C \simeq g^2 
\lim\limits_{p_\mu \rightarrow 0} { g_{0 0} - {p_0 p_0 \over g^2 \kappa} \over p^2 + g^2 \kappa} = {1\over 
\kappa}~,
\eeq 
where we used the perturbative expansion of the action, Eqn.~(\ref{XYaction3}), to calculate the 
zero-momentum two-point function of the massive gauge boson (the overall factor of $g^2$ is due to the field 
normalization and the metric  is Euclidean). Comparing the results of the simulation with the $1/\kappa$ 
curve, also drawn on Fig.~\ref{fig:susc_sin}, we see that  the expected perturbative behavior (\ref{mW}) is 
well reproduced by our Monte-Carlo results for large $\kappa$, and that the Green's function is drastically 
different for small values of $\kappa$.

 The measurements of the scalar susceptibilities,  $\langle \cos A_\mu \cos A_\nu \rangle$, are shown on 
Fig.~\ref{fig:susc_cos}. The states created by $\cos A_{x, 0}$ and $\cos A_{x,1}$  mix, as the $Z_4$ 
symmetry does not forbid them to (in the Hamiltonian formulation, the mixing is between states created by the operators 
Re$(\phi_x^* U_{x, \hat{1}} \phi_{x + \hat{1}})$ and  $  \cos \pi_{x}$). We show the two eigenvalues of the susceptibility matrix on 
Fig.~\ref{fig:susc_cos}.  Clearly, while an 
increase of the susceptibility (and hence the correlation length) around the critical value of $\kappa$ is 
visible on Fig.~\ref{fig:susc_cos}---consistent with the   Pad\'e resummation of the $1/g^2$  expansion 
\cite{Jones:1978yr}---the absence of finite-size scaling indicates that the two-particle scalar states remain 
heavy, with a correlation length of order the lattice spacing.
 
While our main interest in this Section was the susceptibility (the zero-momentum Green's function) and 
its large-volume scaling, we have also measured position-dependent Green's functions. The results 
reveal that, for $\kappa < 1$, the fields are essentially uncorrelated at neighboring lattice sites, and are 
thus heavier than the UV cutoff; we note that this is also the result obtained in \cite{Jones:1978yr}. We do 
not show these data as they are not particularly illuminating; moreover, even if one takes the value of the 
zero-momentum Green's functions for $\kappa \le {\cal{O}}(1)$ as indicative of the square of the Compton 
wavelength of the state (this, in fact, is a gross overestimate, as the true energies of the states go to 
infinity as $\kappa \rightarrow 0$), a quick look at Figs.~\ref{fig:susc_cos}, \ref{fig:susc_sin} shows that 
the maximum Compton wavelength is about $\sqrt{0.7} \approx .84$ in units of the lattice spacing. This is 
consistent with  the strong-coupling expansion in $\kappa$, as described at the end of Section 
\ref{themodel}, as well as with the results of ref.~\cite{Jones:1978yr}, which showed that both  V and S 
states decouple at $\kappa < \kappa_c$.   

\section{Conclusions from the Monte-Carlo study}

We performed a detailed study of a number of observables in the gauged XY-model, for various values of 
$\beta$ and $\kappa$. Here is a summary of the results of this paper:
\begin{enumerate}
\item
We showed that  for $\kappa < \kappa_c$  the string tension   precisely agrees with the exactly known 
pure gauge theory  expression, Eqn.~(\ref{wilsongauge}),  for a number of lattice sizes and values of $
\beta$,  see Figs.~\ref{fig:wilsonareaperimeter8}, \ref{fig:wilsonareaperimeter16}, \ref{fig:wilson8}. This 
even includes observable finite size effects in Fig.~\ref{fig:wilson8}, which correctly reproduce the 
finite-volume pure gauge theory result of Eqn.~(\ref{wilsongaugevolume}).
\item In Fig.~\ref{fig:wilsonkappa}, we observed even the small   ${\cal{O}} (\kappa^4)$  correction to the 
bare gauge coupling of Eqn.~(\ref{XYaction4}) from the leading-order strong-coupling (small-$\kappa$) 
expansion. 
\item We showed that the Polyakov loop expectation value 
  vanishes for  $\kappa$ in the same  range, $\kappa<\kappa_c$, where the string tension is that of the 
pure gauge theory, see Fig.~\ref{fig:polyakovloop}.
\item Our results for  the scalar, Fig.~\ref{fig:susc_cos}, and  vector, Fig.~\ref{fig:susc_sin}, susceptibilities 
(\ref{susc}) revealed no evidence for massless states in the spectrum in the intermediate  regime of $
\kappa \sim \kappa_c$. 
\item Furthermore, the vector susceptibility was shown, see Fig.~\ref{fig:susc_sin}, to match that 
expected of a massive gauge boson for large values of $\kappa \gg 1$, precisely  where perturbation 
theory is expected to be valid, see Section \ref{themodel}. The same zero-momentum propagator is, 
however, drastically different for $\kappa < \kappa_c$, indicating that the correspondent state  is heavier 
than the inverse lattice spacing and decouples from the infrared physics.
\end{enumerate}
These results  show that we have achieved our main objective: to provide numerical evidence for the fact 
that at small  $\kappa < \kappa_c = {\cal{O}}(1)$ the ``gauge symmetry breaking" decouples and that the 
infrared physics of the theory is that of the pure gauge theory, with a small shift of the bare gauge 
coupling. 
\acknowledgments
We thank Joel Giedt for discussions and comments.
We acknowledge support by the National Science and Engineering Research Council of Canada 
(NSERC) and by an Ontario Premier's Research Excellence Award.

\end{document}